\documentclass[preprint,twocolumn]{aastex62}
\usepackage{amsmath,natbib}

\shorttitle{Characteristic Timescales in Blazars}
\shortauthors{Ryan et al.}

\begin{document}

\title{Characteristic Variability Timescales in the Gamma-ray Power Spectra of Blazars}

\correspondingauthor{J. L. Ryan}
\email{jryan@astro.ucla.edu}

\author{J. L. Ryan}
\affiliation{Department of Physics \& Astronomy, University of California, Los Angeles, 430 Portola Plaza, Los Angeles, CA 90095, USA}

\author{A. Siemiginowska}
\affiliation{Harvard-Smithsonian Center for Astrophysics, 60 Garden Street, Cambridge, MA 02138, USA}

\author{M. A. Sobolewska}
\affiliation{Harvard-Smithsonian Center for Astrophysics, 60 Garden Street, Cambridge, MA 02138, USA}

\author{J. Grindlay}
\affiliation{Harvard-Smithsonian Center for Astrophysics, 60 Garden Street, Cambridge, MA 02138, USA}

\begin{abstract}
Characteristic variability timescales in blazar $\gamma$-ray light curves can provide insight into the physical processes responsible for the $\gamma$-ray variability.
The power spectral density (PSD) is capable of revealing such timescales, which may appear as breaks or periodicities.
Continuous-time autoregressive moving-average (CARMA) models can be used to accurately estimate a light curve's PSD.
Through a lightcurve simulation study, we develop a methodology to identify PSD breaks using CARMA models.
Using this methodology, we study the $\gamma$-ray light curves of 13 bright blazars observed with the \textit{Fermi} Large Area Telescope in the 0.1--300 GeV band over 9.5 years.
We present the blazar $\gamma$-ray PSDs, which provide evidence for low-frequency breaks on timescales $\sim$1 year in four sources, and an additional high-frequency break on a timescale $\sim$9 days in one source.
\end{abstract}


\section{Introduction} \label{sec:intro}
% Blazars
Blazars are a subclass of active galactic nuclei (AGN) whose jet is oriented close to our line of sight \citep{ant93,urr95}.
They are characterized by powerful nonthermal emission and rapid, high-amplitude variability across the electromagnetic spectrum, on timescales ranging from minutes to years.
The blazar spectral energy distribution (SED) contains a low-energy component covering radio to X-ray frequencies and a high-energy component peaking in the $\gamma$-rays.
Blazars are further classified as either BL Lac objects (BL Lacs) or flat spectrum radio quasars (FSRQs), based on their optical emission line strength, and lie on opposing ends of the ``blazar sequence'' \citep{fos98,ghi17}.

%Problem / variability
While the blazar jet is believed to be powered by accretion onto a black hole \citep{bla74}, the exact mechanisms responsible for the emission and acceleration are not fully understood.
Characterizing the blazar variability at multiple wavelengths is one important way of constraining physical models.
Of particular interest is the $\gamma$-ray regime, where the high-energy component of the SED peaks, and which has been studied in less detail than the radio, optical, and X-ray bands.
The \textit{Fermi} Large Area Telescope (\textit{Fermi}/LAT) blazar monitoring program \citep{abd10c} has provided an excellent opportunity to study the long-term $\gamma$-ray variability.

% Solution / The PSD
One way of quantifying variability is with the power spectral density (PSD), which describes the amplitude of variations in a time series as a function of Fourier frequency (or variability timescale).
The PSD is useful since it reveals average properties of the variability, whereas the light curve may be thought of as only a single realization of an underlying stochastic process \citep[e.g.][]{vau03}.
Additionally, other methods of variability analysis, such as multi-wavelength correlation studies, rely on accurate PSD models to assess the significance of their results \citep{cha08}.
Periodicities will also appear in the PSD if they are present in the data.

% History / CARMA
While previous studies have characterized the $\gamma$-ray PSDs of blazars \citep[e.g.][]{abd10c,ack10,nak13,sob14,ram15,kus17}, limitations were imposed by both methodology and light curve length.
Fourier-based methods, as well as the Lomb-Scargle periodogram \citep{lom76,sca82}, produce a PSD estimate subject to distortions from aliasing and red-noise leak \citep{dee75,vau03,kel11}.
Parametric methods can address these issues \citep{don92,utt02,kel09}, but require a suitably flexible PSD form to be chosen.
A simple power law, for example, may not adequately describe the $\gamma$-ray PSD of the blazar 3C 454.3, which has shown evidence for a change in PSD slope on a timescale of a few days \citep{ack10,nak13}.
Frequencies where the PSD slope changes, or ``breaks,'' may be indicative of characteristic timescales associated with variability.

% CARMA
In the present work we aim to produce more accurate estimates of blazar PSDs using continuous-time autoregressive moving average (CARMA) model fitting, through the \texttt{carma\_pack}\footnote{\url{http://github.com/brandonckelly/carma_pack.}} implementation of \citet{kel14}.
The technique assumes that a light curve is the realization of a Gaussian process, and uses Bayesian inference to calculate the light curve's model parameters, which in turn are used to directly compute the underlying PSD.
CARMA models generalize the CAR(1) model \citep[equivalently the damped random walk, or Ornstein-Uhlenbeck process;][]{kel09} to higher order, and also have the flexibility to account for multiple break-like features and quasi-periodic oscillations (QPOs) in their PSDs.
We note that the PSDs of Gaussian processes not equivalent to CARMA models have also been explored \citep[e.g.][]{ras05,wil13,for17}.

% What we do
We fit CARMA models to the $\gamma$-ray light curves of 13 bright blazars observed by \textit{Fermi}/LAT.
We conduct a light curve simulation study to help interpret the CARMA model results.
Our sample includes 8 FSRQs and 5 BL Lacs in order to look for potential differences between the two classes.
The $\sim$9.5 year light curves allow for the study of lower Fourier frequencies than previously.
After modeling the PSD, we report the break frequencies (if any) and average PSD slopes.

% Outline
This paper is organized as follows.
In \S \ref{sec:data} we describe the data.
In \S \ref{sec:analysis} we describe CARMA models, explore the limits of CARMA modeling through simulations, and describe our methodology.
In \S \ref{sec:result} we report the results of our CARMA model fits, and present the derived PSDs.
In \S \ref{sec:disc} we discuss our results and potential physical interpretations.
In \S \ref{sec:conc} we summarize our conclusions.

%TABLE 1
\newcommand{\tableonecomment}{
For each source, results are reported for the analyses of both the daily (above) and weekly (below) light curves.
SED type (either LSP, ISP, or HSP, corresponding to low, intermediate, or high synchrotron peak) and redshift are obtained from 3LAC \citep{ack15a} unless otherwise noted.
Black hole masses are averages of the values in \citet{ghi09b}, \citet{xio14}, \citet{kra16}, and references therein, with the root-mean-squares of the values used as uncertainties.
Long ($\tau_L$) and short ($\tau_S$) characteristic timescales are reported in the observer's frame.
Average slopes are obtained by fitting power laws to the portion of the PSD above the estimated noise level.
Uncertainties on timescales and average slopes represent 90\% CIs, while limits represent 99\% CIs.
An asterisk indicates that the timescale is determined to be artificial, as described in \S \ref{sec:car_method}.
The column labeled $p$ contains the CARMA($p,p-1$) model order chosen by the model selection procedure.
The $\chi^2$/d.o.f. column reports the $\chi^2$, the sum of squared standardized residuals between the light curve and best-fit CARMA process, and the degrees of freedom, the number of data points minus the number of model parameters.
}
\begin{deluxetable*}{lcccccccc}[!ht]
\tablecolumns{9}
\tablewidth{\skip54}
\tablecaption{List of Analyzed Blazars and Results of PSD Fitting}
\tablehead{\colhead{Source Name} & \colhead{SED Class} & \colhead{Redshift} & \colhead{$\log (M_{BH}/M_\odot )$} & \colhead{$\log  (\tau_L$/day)} & \colhead{$\log  (\tau_S$/day)} & \colhead{Avg. Slope} & \colhead{$p$} & \colhead{$\chi^2/$d.o.f.} }
\startdata
\multicolumn{9}{c}{FSRQs} \\ \tableline
B2 1633+38 & LSP & 1.814 & $9.43 \pm 0.25$ & $> 2.47$ & \;$0.81^{+0.20}_{-0.15}$* & $-1.24^{+0.19}_{-0.23}$ & 2 & 595.5/730 \\
& & & & $> 2.57$ &  & $-1.31^{+0.26}_{-0.66}$ & 1 & 325.9/359 \\
PKS 1424-41 & LSP & 1.522 & $9.00 \pm 1.50$ & $2.57^{+0.36}_{-0.15}$ & $< 0.72$ & $-1.22^{+0.18}_{-0.24}$ & 2 & 1277.4/1458 \\
& & & & $2.64^{+0.36}_{-0.11}$ &  & $-1.30^{+0.21}_{-0.31}$ & 1 & 406.5/402 \\
B2 1520+31 & LSP & 1.487 & $9.16 \pm 0.24$ & $> 2.75$ & \;$1.17^{+0.17}_{-0.19}$* & $-0.99^{+0.29}_{-0.43}$ & 2 & 462.2/542 \\
& & & & $> 3.09$ & \;$1.83^{+0.10}_{-0.22}$* & $-1.14^{+0.19}_{-0.22}$ & 2 & 353.0/372 \\
PKS 0454-234 & LSP & 1.003 & $9.40$ & $2.34^{+0.46}_{-0.24}$ & $< 0.79$ & $-0.88^{+0.21}_{-0.26}$ & 2 & 629.8/719 \\
& & & & $2.51^{+0.30}_{-0.11}$ &  & $-1.01^{+0.19}_{-0.30}$ & 1 & 372.3/367 \\
3C 454.3 & LSP & 0.859 & $8.82 \pm 0.22$ & $> 2.46$ & \;$1.19^{+0.16}_{-0.20}$* & $-1.33^{+0.15}_{-0.20}$ & 2 & 1931.6/2215 \\
& & & & $> 2.51$ &  & $-1.44^{+0.19}_{-0.26}$ & 1 & 409.3/420 \\
3C 279 & LSP & 0.536 & $8.50 \pm 0.24$ & \;$2.04^{+0.47}_{-0.32}$* & $0.98^{+0.18}_{-0.32}$* & $-0.74^{+0.13}_{-0.18}$ & 2 & 1021.3/1147 \\
& & & & \;$2.28^{+0.17}_{-0.11}$* &  & $-0.89^{+0.13}_{-0.17}$ & 1 & 398.5/416 \\
PKS 1510-089 & LSP & 0.360 & $8.36 \pm 0.27$ & $2.29^{+0.29}_{-0.17}$ & $0.94^{+0.13}_{-0.15}$ & $-0.90^{+0.11}_{-0.14}$ & 2 & 1121.4/1237 \\
& & & & $2.27^{+0.17}_{-0.08}$ &  & $-0.93^{+0.12}_{-0.15}$ & 1 & 436.9/445 \\
3C 273 & LSP & 0.158 & $8.51 \pm 0.75$ & $> 2.64$ & \;$1.09^{+0.12}_{-0.20}$* & $-1.03^{+0.19}_{-0.21}$ & 2 & 287.1/331 \\
& & & & $> 2.33$ & $< 1.99$ & $-1.12^{+0.26}_{-0.29}$ & 2 & 235.6/239 \\
\tableline
\multicolumn{9}{c}{BL Lacs} \\ \tableline
3C 66A & ISP & \;0.444\tablenotemark{a} & $8.30 \pm 0.30$ & $> 2.67$ & \;$1.08^{+0.18}_{-0.45}$* & $-1.42^{+0.53}_{-0.25}$ & 2 & 257.7/332 \\
& & & & $> 2.76$ & $< 1.85$ & $-1.20^{+0.35}_{-0.37}$ & 2 & 354.0/420 \\
PKS 0716+714 & ISP & \;0.127\tablenotemark{b} & $8.10$ & $> 2.01$ & \;$0.92^{+0.19}_{-0.14}$* & $-0.88^{+0.29}_{-0.29}$ & 2 & 864.2/1098 \\
& & & & \;$2.22^{+0.17}_{-0.10}$* &  & $-0.67^{+0.12}_{-0.17}$ & 1 & 411.2/426 \\
PKS 2155-304 & HSP & 0.116 & $8.70$ & $> 2.14$ & $< 1.18$ & $-0.85^{+0.41}_{-0.64}$ & 2 & 578.8/733 \\
& & & & \;$2.21^{+0.23}_{-0.13}$* &  & $-0.43^{+0.14}_{-0.20}$ & 1 & 436.1/436 \\
BL Lac & ISP & 0.069 & $8.21 \pm 0.41$ & $> 2.24$ & \;$0.80^{+0.13}_{-0.17}$* & $-0.85^{+0.25}_{-0.33}$ & 2 & 671.0/752 \\
& & & & $> 1.97$ & $< 1.96$ & $-0.95^{+0.32}_{-0.39}$ & 2 & 388.0/333 \\
Mkn 421 & HSP & 0.031 & $8.15 \pm 0.34$ & $2.69^{+0.40}_{-0.23}$ & $< 0.77$ & $-0.96^{+0.29}_{-0.46}$ & 2 & 1158.5/1649 \\
& & & & $2.59^{+0.36}_{-0.14}$ &  & $-0.89^{+0.25}_{-0.42}$ & 1 & 526.2/485 \\
\enddata
\tablenotetext{a}{\citet{tor18} report a redshift of 0.340.}
\tablenotetext{b}{\citet{nil08} and \citet{dan13} report a redshift $\sim$0.3.}
\tablecomments{\tableonecomment}
\label{table:results}
\end{deluxetable*}

\section{Data} \label{sec:data}
We study 13 bright blazars that include 8 FSRQs and 5 BL Lac objects, observed by \textit{Fermi}/LAT.
The sample was previously analyzed by \citet{sob14}, and was chosen here for the purpose of comparison.
We use the Monitored Source List Light Curves provided by the \textit{Fermi} Science Support Center\footnote{\label{footnote2}\url{http://fermi.gsfc.nasa.gov/ssc/data/access/lat/msl_lc/}}, retrieved on 2018 January 22.
We analyze both the weekly and daily binned light curves provided to assess biases (if any) introduced by the choice of the time bin size.
The light curves include observations made between MJD 54684--58139 (2008 August 6 to 2018 January 21) in the energy range of 0.1--300 GeV.
\citet{sob14} used 4 year-long light curves and an adaptive binning scheme based on the signal-to-noise ratio (S/N).

The source names and other properties are listed in Table \ref{table:results}.
SED type (either LSP, ISP, or HSP, corresponding to low, intermediate, or high synchrotron peak) and redshift are obtained from 3LAC \citep{ack15a}.
Black hole masses are averages of values reported in \citet{ghi09b}, \citet{xio14}, \citet{kra16}, and references therein, and the uncertainties are the root-mean-squares of the values.
For PKS 1424-41, we use the mass estimate of \citet{kra16} with the parameters of \citet{bon13}.

The light curves do not contain detections for every day or week, and in some cases flux upper limits are provided.
As described on the \textit{Fermi} Science Support Center website\textsuperscript{\ref{footnote2}}, the data are not absolute flux measurements, and are produced using preliminary instrument response functions and calibrations.

\section{Analysis} \label{sec:analysis}

We describe CARMA models and the statistics used to evaluate them in \S \ref{sec:car_models}.
In \S \ref{sec:car_sims}, we perform a lightcurve simulation study to evaluate the limits of CARMA modeling, with emphasis on break timescale recovery.
Using the simulation results, we develop a methodology to identify PSD breaks using CARMA models in \S \ref{sec:car_method}.

\subsection{CARMA Models} \label{sec:car_models}
% Kelly+, 2014
We estimate the underlying PSDs of the blazar $\gamma$-ray light curves using the CARMA model fitting technique of \citet{kel14}.
CARMA models handle irregular sampling and measurement uncertainties in the light curves by fitting the data in the time domain, and their PSDs have a flexible parametric form.
Like \citet{sob14} we model the natural logarithm of the \textit{Fermi} light curves, as the light curve flux distributions are closer to lognormal than normal \citep{kus16}, following considerations put forward by \citet{utt05}.
Since the flux upper limits in our data are poorly approximated by data points with Gaussian errors in log-space and \texttt{carma\_pack} only handles Gaussian errors, we exclude upper limits from our analysis and explore the effects of doing so in our light curve simulations.

% What is CARMA?
Each light curve is modeled as a CARMA process $y(t)$, defined as the solution to
\begin{equation}\label{eqn:car_y}
\begin{split}
\frac{d^p y(t)}{dt^p}+\alpha_{p-1}\frac{d^{p-1} y(t)}{dt^{p-1}}+\dots+\alpha_0 y(t) \qquad\quad \\
=\beta_q \frac{d^q \epsilon(t)}{dt^q}+\beta_{q-1}\frac{d^{q-1} \epsilon(t)}{dt^{q-1}}+\dots+\epsilon(t)
\end{split}
\end{equation}
where $\epsilon(t)$ is a Gaussian noise process with variance $\sigma^2$ and mean zero.
The order of the autoregressive polynomial is $p$, the order of the moving average polynomial is $q$, and the corresponding CARMA model is notated as CARMA($p,q$).
The free parameters in the model are the autoregressive coefficients $\vec{\alpha}=(\alpha_0,\dots,\alpha_p)$, the moving-average coefficients $\vec{\beta}=(\beta_0,\dots,\beta_q)$, and $\sigma$, with $\alpha_p=\beta_0=1$.
The PSD associated with a CARMA($p,q$) process is given analytically by 
\begin{equation}\label{eqn:car_psd}
P(f)=\sigma^2\frac{|\sum_{j=0}^q\beta_j(2\pi i f)^j|^2}{|\sum_{k=0}^p\alpha_k(2\pi i f)^k|^2}
\end{equation}
and can be expressed as the sum of $p$ Lorentzians \citep{kel14}.

% Breaks
For Lorentzians with centroids of 0, their widths correspond to break frequencies, where the PSD steepens toward higher frequencies.
In the CARMA(1,0) model, the PSD transitions from flat to $P(f)\propto f^{-2}$ at higher frequencies, while higher-order models steepen by some integer multiple of 2, in general.
Lorentzians with non-zero centroids may represent QPOs.
Both features indicate the presence of characteristic timescales associated with variability.

% params, MCMC
For each CARMA($p,q$) model, the parameters $\vec \alpha$, $\vec \beta$, and $\sigma$, as well as a scaling parameter on the measurement errors, are sampled using Markov Chain Monte Carlo (MCMC) methods to find their maximum-likelihood estimates (MLE) and to derive confidence intervals (CIs).
A parallel tempering algorithm is used to attempt to account for multi-modal likelihoods, common when $p>1$ \citep{kel14}.
Throughout this work, we run the MCMC sampler for $5\times 10^4$ iterations after a burn-in of $2.5\times 10^4$ iterations.

% norm
We normalize the CARMA PSD using the factor
\begin{equation}
A=2\Delta T_\text{samp}/N
\end{equation}
where $\Delta T_\text{samp}$ is the median sampling time and $N$ is the number of data points in the light curve. This normalization gives the periodogram in absolute units (e.g. (ct cm$^{-2}$ s$^{-1})^2$ / day$^{-1}$), and the integral of the PSD is equal to the variance of the light curve \citep{vau03}.
Under this normalization, assuming Gaussian errors, the noise level is given by 
\begin{equation}\label{eqn:noise}
P_\text{noise}=2\Delta T_\text{samp}\overline{\sigma_\text{err}^2}
\end{equation}
where $\overline{\sigma_\text{err}^2}$ is the unweighted mean square light curve measurement error.
The noise level treats measurement errors as a white noise process, and thus represents a constant that would be added to the underlying light curve's PSD.
Since the measurement error variance is not constant, and also exhibits flux dependence, the noise level is only approximate.

% DIC
To perform model selection, we evaluate our models using the Deviance Information Criterion \citep[DIC;][]{spi02} and the Ljung-Box test \citep{lju78}.
The DIC is defined as
\begin{equation}\label{eqn:dic}
\text{DIC}=\bar D+p_D
\end{equation}
where $\bar D$ is the mean of the deviance (equal to $-2\;\times$ the log-likelihood of a set of parameters) and $p_D$ is the effective number of parameters.
The DIC thus decreases with increasing goodness of fit, and increases with model complexity.
In our analysis, and in the implementation of \citet{kel14}, $p_V=\text{var}(D)/2$ is used in place of $p_D$, as detailed by \citet{spi14}.

% LB
Additionally, we require that a model adequately captures the correlation structure in a light curve using the Ljung-Box test.
The sequence of standardized residuals of an adequate CARMA model fit to a time series should have the properties of Gaussian white noise.
The Ljung-Box test is a procedure to assess the statistical significance of apparent departures of the autocorrelations of these residuals from zero \citep{box94}.
For an appropriate CARMA($p,q$) model with residual autocorrelations $r_k$, the modified Ljung-Box-Pierce statistic
\begin{equation}\label{eqn:car_ljungbox}
Q=n(n+2)\sum_{k=1}^m (n-k)^{-1}r_k^2
\end{equation}
is distributed as $\chi^2_{m-p-q}$ for large $n$, where $n$ is the number of light curve data points, and $n$ is large relative to $m$ \citep{lju78}.
We can therefore calculate the probability (i.e. the p-value) of observing values at least as large as $Q$ given its expected distribution.
Large values of $Q$ suggest that the autocorrelations are greater than expected, given residuals sampled from an independent, identically distributed sequence, resulting in a small p-value.
As the formula makes evident, the Ljung-Box test does not include autocorrelations longer than $m$ lags, so we would like to choose $m$ as large as possible while maintaining the validity of the test.
We use $m=\lfloor0.05\times n\rfloor$, where $\lfloor x \rfloor$ is the greatest integer less than or equal to $x$.
If this results in $m-p-q<1$, we instead use $m=p+q+1$.

% preliminary model sel / ps considered
Our modeling procedure for a given light curve consists of fitting CARMA($p,p-1$) models for $1 \leq p \leq 5$, then performing model selection.
Model selection proceeds by first excluding all CARMA models whose Ljung-Box p-value is $<0.01$, then choosing the lowest-order model whose DIC is within 10 of the minimum DIC among all models, to avoid overfitting.
We also require the model to have a $\chi^2$ test \citep[e.g.][]{pre07} p-value $<0.01$.
We investigate only CARMA($p,p-1$) models since they can approximate any discretely-sampled, stationary stochastic system with arbitrary precision \citep{pan83}.
We consider only models with $1 \leq p \leq 5$, as we find that the DIC does not significantly improve past $p=2$ in our modeling of the data in \S \ref{sec:result}.

\begin{figure}[t!]
  \plotone{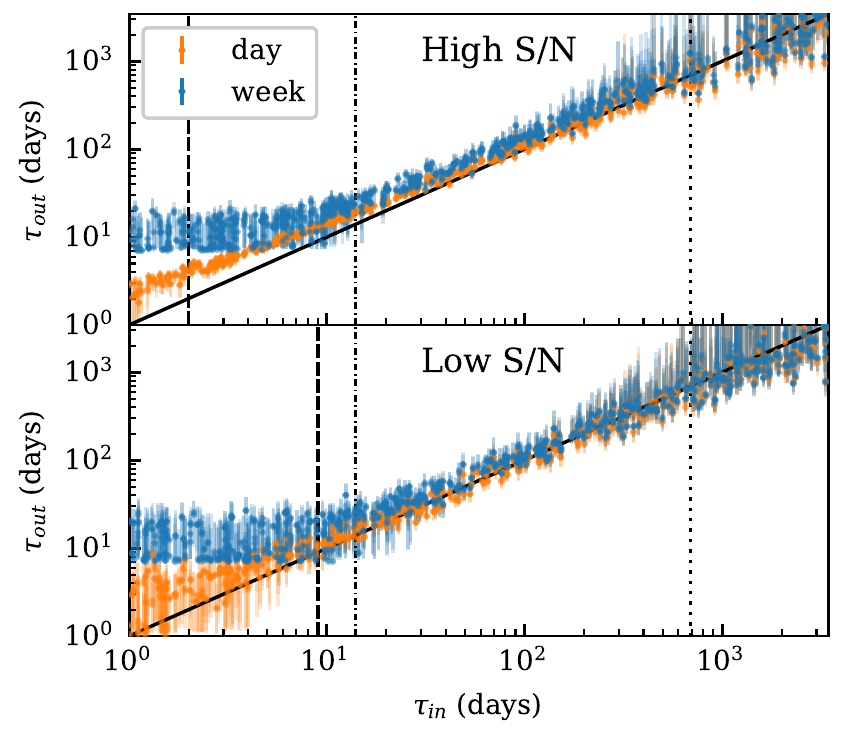}
  \caption{Recovered timescales $\tau_\text{out}$ from CARMA(1,0) simulations of daily-binned (orange) and weekly-binned (blue) light curves, as a function of the true input timescales $\tau_\text{in}$.
Error bars represent 90\% CIs.
The upper plot corresponds to the high-S/N simulations, while the lower plots correspond to the low-S/N simulations.
The solid black line represents $\tau_\text{out}=\tau_\text{in}$.
The $\tau_\text{out}$ are systematically longer than $\tau_\text{in}$ due to binning effects; CARMA(2,1) models are required for accurate parameter recovery, as discussed in \S \ref{sec:car_sims}.
The true timescale is recovered consistently only when $\tau_\text{in} > 2$ or 9 days (dashed lines) in the high-S/N and low S/N daily simulations, respectively, $\tau_\text{in} > 14$ d (dash-dotted lines) in the weekly simulations, and $\tau_\text{in} \lesssim 0.2T_\text{exp}$, 20\% of the experiment length (dotted lines).
  \label{fig:koz}}
\end{figure}

\subsection{CARMA Simulations} \label{sec:car_sims}
% purposes
Limitations on PSD parameter recovery, as well as spurious identifications of variability features such as break timescales or QPOs, can be present in variability analyses \citep[e.g][]{emm10,vau16,koz16,koz17}.
We therefore perform simulations to evaluate how accurately our CARMA modeling methodology recovers a light curve's PSD, given the quality of our data.

% LC generation
We use the algorithm of \citet{tim95} to generate light curves from an input PSD.
Since the recovery of CARMA(1,0) model parameters depends on light curve length, cadence, and measurement error properties \citep{koz17}, we degrade the simulated light curves to match the properties of the data, including the light curves' means, variances, lengths, and error versus flux distributions.
We generate light curves of length 3455 days with a 3 hour cadence, the time it takes \textit{Fermi} to view the entire sky \citep{atw09}.
The light curves are scaled and offset to match the data mean and variance, then binned into days using the mean of the the data in each bin.
For each daily-binned flux value, an uncertainty is drawn from the empirical distribution of measurement errors with fluxes within 0.05 dex of the simulated daily-binned flux, and Gaussian white noise with a corresponding amplitude is added.
Following \citet{abd10c}, we consider fluxes to be upper limits when their S/N $\leq 2$.
Days for which the data include neither a detection nor an upper limit are also omitted.
If the simulated light curve still contains more flux detections than the data, a number of days equal to the difference are randomly selected and omitted.
The daily light curves are binned into week-long bins, using weighted averages.
We degrade the simulated light curves to the data quality of either 3C 454.3 or BL Lac, chosen due to their difference in light curve quality---their mean S/N is 7.6 and 5.1 respectively, and consequently the fraction of data points considered non-detections/upper limits is higher for BL Lac (36\% for 3C 454.3 and 77\% for BL Lac).

% CAR(1) / Kozlowski17 test
We first apply our modeling procedure to light curves with PSDs containing a break, to understand the limits on which timescales we can recover.
We simulate 1000 light curves with CARMA(1,0) PSDs whose decorrelation timescales lie between 1 day and the length of our data (3455 days); i.e. the input PSDs are of the form $P(f)\propto1/(f^2+\tau_\text{in}^{-2})$ with 1 day $<$ $\tau_\text{in}$ $<$ 3455 days.

% CAR(1) / Kozlowski17 test - results
We find that the break timescales can be recovered from both daily and weekly versions of both the high-S/N and low-S/N light curves only for a certain range of timescales.
The CARMA(1,0) model is usually selected as the best-fit model (i.e. is the lowest-order model which has a DIC within 10 of the minimum, and has a Ljung-Box p-value $>0.01$), while the CARMA(2,1) model is selected in a small percentage ($\sim 2$\%) of the simulations.
Plots of the output CARMA(1,0) model timescales $\tau_\text{out}$ and their 90\% CIs, as a function of $\tau_\text{in}$, are shown in Figure \ref{fig:koz}.
In the weekly-binned simulations, the true timescale is recovered consistently only when $\tau_\text{in} > 14$ days and $\tau_\text{in} \lesssim 0.2T_\text{exp}$, 20\% of the experiment length.
This is similar to the result of \citet{koz17}, where it is reported that the light curve must be at least ten times longer than $\tau_\text{in}$, in order to recover $\tau_\text{out}$ from SDSS \citep{yor00} or OGLE-III \citep{uda15} light curves.
In the daily-binned simulations, $\tau_\text{in}$ must be greater than 2 days for high-S/N data, and greater than 9 days for low-S/N data.
The minimum recoverable $\tau_\text{in}$ from the low-S/N data is due to the noise level, rather than corresponding to the Nyquist frequency.
The ability to recover breaks near the noise level is explored in the next set of simulations.
These approximate limits, beyond which the $\tau_\text{out}$ versus $\tau_\text{in}$ curve flattens, are found by moving outward from 100 days and finding the first 1-day wide $\tau_\text{out}$ bin where the $\tau_\text{in}$ distribution includes values less than 2 days or greater than 3455/2 days in the 90\% CI.

\begin{figure}[t]
  \plotone{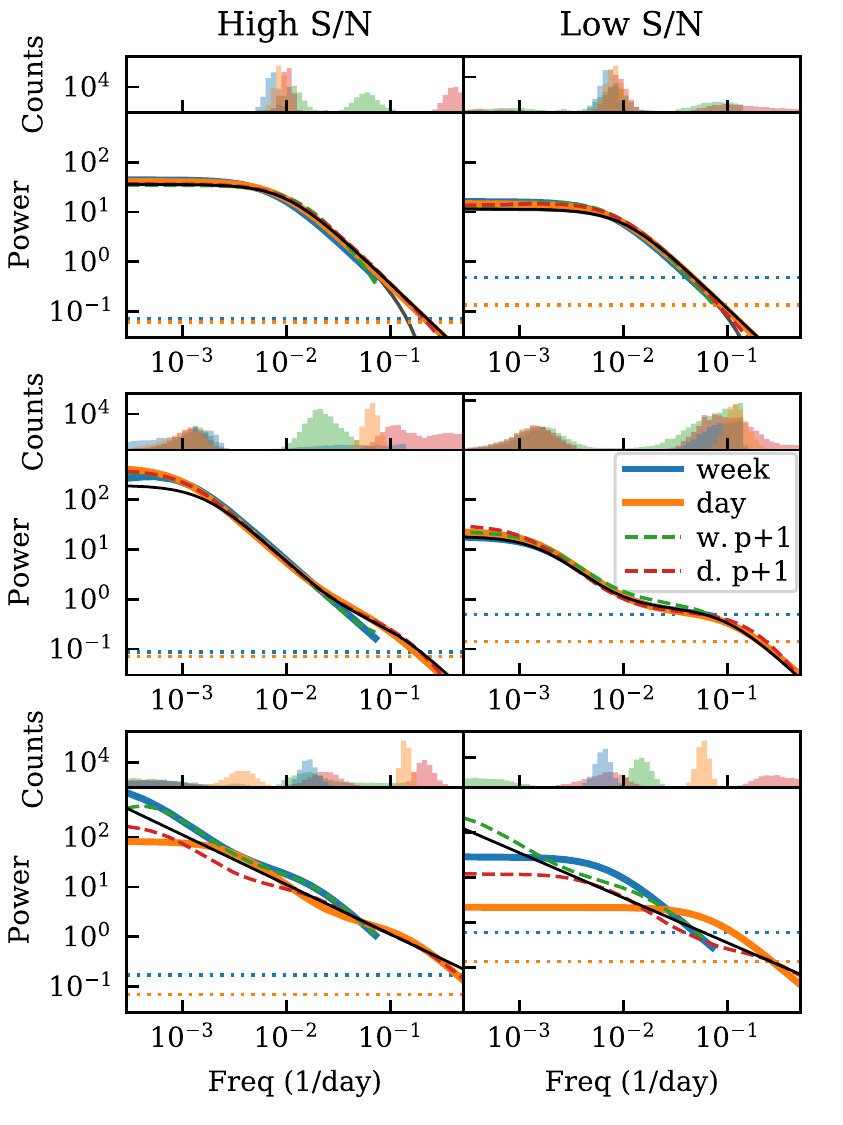}
  \caption{CARMA model PSDs and break histograms summarizing the simulation results.
From top to bottom, the true PSDs have the form of a CARMA(1,0) PSD with a break timescale of 100 days, a CARMA(2,1) PSD with breaks at 10 and 600 days, and $1/f$ noise.
The left plots correspond to the high-S/N simulations, while the right plots correspond to the low-S/N simulations.
Power is in units proportional to log(ct cm$^{-2}$ s$^{-1})^2$ / day$^{-1}$.
In the PSD plots, the true PSD (solid black line), along with the daily (solid orange line) and weekly (solid blue line) CARMA model MLE PSDs are shown.
The MLE PSDs of the CARMA models of the next highest order for the daily (dashed red line) and weekly (dashed green line) light curves are also shown.
The solid gray line in the top-left plot shows the effect of 7-day binning on the true PSD.
The horizontal orange and blue dotted lines represent the noise levels for the daily and weekly PSDs, respectively.
Above the PSD plots, histograms of the entire MCMC sample of CARMA model break timescales are plotted, in colors matching the PSDs.
The top and middle plots illustrate that PSD breaks, when present, are recovered in the best-fit models and the next higher-order models.
The bottom plots illustrate that CARMA modeling can introduce artificial breaks in non-CARMA PSDs, as for a power-law PSD with index $\alpha=-1$, but the breaks do not show agreement between the selected models and the next higher-order models, or between daily and weekly PSDs.
  \label{fig:sims}}
\end{figure}

% CAR(1) / Kozlowski17 test - systematically high
We also find that the break timescales recovered from the CARMA(1,0) models are systematically longer than the true timescales, particularly in the high-S/N simulations, due to an effect of binning the PSD.
For the high-S/N simulations, the mean $\tau_{out}/\tau_{in}$ ratios are 1.2 (daily) and 1.3 (weekly), and for the low-S/N simulations, the medians are 1.0 (daily) and 1.1 (weekly), all with standard deviations of $0.2$.
In contrast, the $\tau_{out}/\tau_{in}$ ratios from CARMA(2,1) models are all $1.0 \pm 0.1$.
When the data is binned, the light curve is effectively convolved with a rectangle function of width $T_\text{bin}$.
The PSD is thus multiplied by the rectangle function's PSD, $P(f) \propto \sin^2 (\pi f T_\text{bin})/f^2$.
This effect only becomes noticeable at frequencies close to $1/T_\text{bin}$, so it is negligible when $T_\text{bin}$ is small compared to the sampling cadence, $T_\text{samp}$.
However, $T_\text{bin}$ is equal to $T_\text{samp}$ in our data.
This distorts the PSD by steepening it at higher frequencies, causing the MLE CARMA(1,0) PSD break timescale to appear at lower frequencies so that there is less power at higher frequencies.
However, we find that CARMA(2,1) models recover the break timescales without systematic offsets, by including a second break to $P(f)\propto f^{-4}$ above a frequency $\sim 1/T_\text{samp}$.
This effect is illustrated in the top panels of Figure \ref{fig:sims}.

% CARMA(2,1) test
We next apply our modeling procedure to light curves with CARMA(2,1) PSDs, to understand the limits on recovering higher-frequency breaks near the noise level.
We simulate light curves with CARMA(2,1) PSDs with a low-frequency break at 600 days, and a high frequency break at 10 days.
We vary the amplitude of the Lorentzian producing the high frequency break, and seek to determine the ratio of the amplitude to the estimated noise level below which the high frequency break is not recovered.
We expect to constrain the high-frequency break only with the daily-binned light curves.

% CARMA(2,1) test - results
The simulations reveal that when the high-frequency Lorentzian's amplitude is at least twice the noise level, a CARMA(2,1) model is selected and the 10 day break is recovered in both the high and low-S/N daily-binned light curve simulations.
In the weekly-binned light curves, the 10 day break is usually unconstrained, although when the amplitude of the high-frequency component is low relative to the low-frequency component, a CARMA(1,0) model is selected and the break is not seen, as shown in Figure \ref{fig:sims}.
The 600 day break is recovered in all cases.
We note that the $\chi^2$ values are quite low in many of the CARMA(2,1) models, suggestive of overfitting, despite the correct model being chosen.
This implies that the CARMA model is fitting some of the variability introduced by measurement errors, although this does not significantly distort the PSD.
We therefore do not reject models on the basis of low $\chi^2$.
In many of these simulations, although the CARMA(2,1) model does not improve the DIC, it passes the Ljung-Box test while the CARMA(1,0) model often does not, highlighting the importance of the Ljung-Box test.
When PSD breaks occur below twice the noise level, we recover the break in only $\sim 10\%$ of the simulations.
CARMA(1,0) models begin to be selected, and the high frequency break in the CARMA(2,1) models tend to be either offset from the true break frequency or unconstrained.
We note that this level is only approximate, and furthermore may depend on all the CARMA(2,1) parameters, whereas only the moving average coefficient was varied in these simulations.

% figure - all sims

% PL test
Lastly, we simulate light curves with power law PSDs to search for spurious PSD features and identify ways to recognize them as such.
We simulate 1000 light curves with PSDs of the form $P(f) \propto f^{\alpha}$, with $\alpha$ uniformly randomly distributed between $-4$ and $0$. We include power laws as steep as $\alpha=-4$ since power-law slopes steeper than $-2$ have been reported for optical PSDs of AGN \citep[e.g.][]{mus11}.

% PL test - results
The simulations reveal that spurious breaks can appear for certain PSD power-law slopes, but that such breaks are distinguishable from real ones.
PSDs with $\alpha$ close to 0 or $-2$ are limits of the parametric form of the CARMA(1,0) PSD with sufficiently long or short break timescales, respectively.
Consequently no spurious, constrained breaks arise in the simulations with these slopes.
The simulations with $\alpha\approx-4$ were free of spurious breaks in their PSDs as well, but require CARMA(2,1) models to produce $\alpha=-4$ PSDs.
The remaining power law simulations contain spurious breaks, in general.
This is illustrated in the bottom panels of Figure \ref{fig:sims}, which show representative PSDs from a single $\alpha=-1$ simulation.
Unlike the real breaks in the CARMA(1,0) simulations, which exist near the same frequency in all CARMA models of both daily and weekly light curves, the spurious breaks change location in higher-order models as the CARMA PSD continues to better approximate a power law.
Spurious breaks can be identified as those whose break frequency's 90\% CI does not contain a break's MLE frequency in the next higher-order CARMA model.
Models in which the weekly PSD introduces a constrained break not present in any of the daily light curve's break frequency 90\% CIs also indicate unphysical breaks.

% PL test - results
CARMA(2,1) models are selected for most power-law PSD light curve simulations where $\alpha$ is not an integer multiple of 2.
We find that CARMA(2,1) models pass the Ljung-Box test, indicating that they sufficiently capture the correlation structure in the light curves, whereas the CARMA(1,0) models do not.
Additionally, the CARMA(2,1) models provide significantly better fits to both the light curve and PSD than the CARMA(1,0) model: for a typical high-S/N light curve simulation with an $\alpha=-1$ PSD, the CARMA(2,1) model improves the $\chi^2$ by around 200, and the mean squared error of the estimated break location decreases by $\sim$80\%.
We also find that the PSD slope can be recovered by calculating the average power-law slope of the CARMA model PSD above the noise level.
This is done by performing a least-squares fit of a straight line to each MCMC sample PSD in log-log space, allowing for both the MLE slope and a 90\% CI to be calculated.

\subsection{Methodology} \label{sec:car_method}
Our CARMA modeling methodology which we apply to the \textit{Fermi} light curves is as follows:
\begin{enumerate}
\item Fit CARMA($p,p-1$) models for $1 \leq p \leq 5$.
\item Perform model selection, choosing as the best-fit model the lowest-order CARMA model with Ljung-Box p-value $>0.01$, $\chi^2$ test p-value $>0.01$, and DIC within 10 of the minimum DIC.
\item Within the best-fit model, identify the widths of Lorentzian PSD components with centroids of 0 as potential break frequencies. A break is determined to be unphysical if its 90\% CI does not contain a MLE break in the next higher-order CARMA model, or if the weekly PSD contains a break inconsistent with the 90\% CIs in the daily PSD.
\end{enumerate}

Artificial breaks can arise from the parametric form of CARMA models, since the simplest model we consider, CARMA(1,0), contains a break.
CARMA models of a light curve with a featureless power-law PSD will contain constrained breaks if the power law is not a limiting case of a CARMA model PSD, in general.
However, the break frequencies of CARMA models of progressively higher order do not converge, as they do when the true PSD contains a break.
These breaks also change location between daily and weekly models.
We therefore reject breaks whose frequencies change significantly in the next higher-order CARMA model, or change between the daily and weekly PSD, as this behavior indicates the PSD may be a featureless power law.
We find that we are able to recognize power-law PSDs and estimate their slopes, despite not explicitly including a power-law PSD model.
Processes with power-law PSDs solve a different stochastic differential equation than Eq. \ref{eqn:car_y} \citep[e.g.][]{kas95}, the inclusion of which is beyond the scope of our analysis.
Alternatively, as \citet{kas95} mentions, power laws with slopes between 0 and $-2$ can be approximated as ``superpositions of Lorentzian spectra,'' an approach similar to sup-OU modeling \citep{kel11,sob14}, but with break timescales fixed to values outside the observed frequency range.

Comparing our methodology to \citet{sob14}, we see that our simplest models, the CARMA(1,0) and the OU process, are identical.
Our analyses diverge in their use of sup-OU models, versus our use of CARMA($p,p-1$) models.
While the sup-OU models can more exactly approximate power-law PSDs, CARMA models can account for more general PSD shapes, such as ones including additional breaks, steeper PSD slopes, or QPOs.
We expect our additional ability to explicitly identify artificial breaks and model the measurement errors to provide PSD estimates closer to the true PSDs.

\section{Results} \label{sec:result}
\begin{figure}[t]
  \plotone{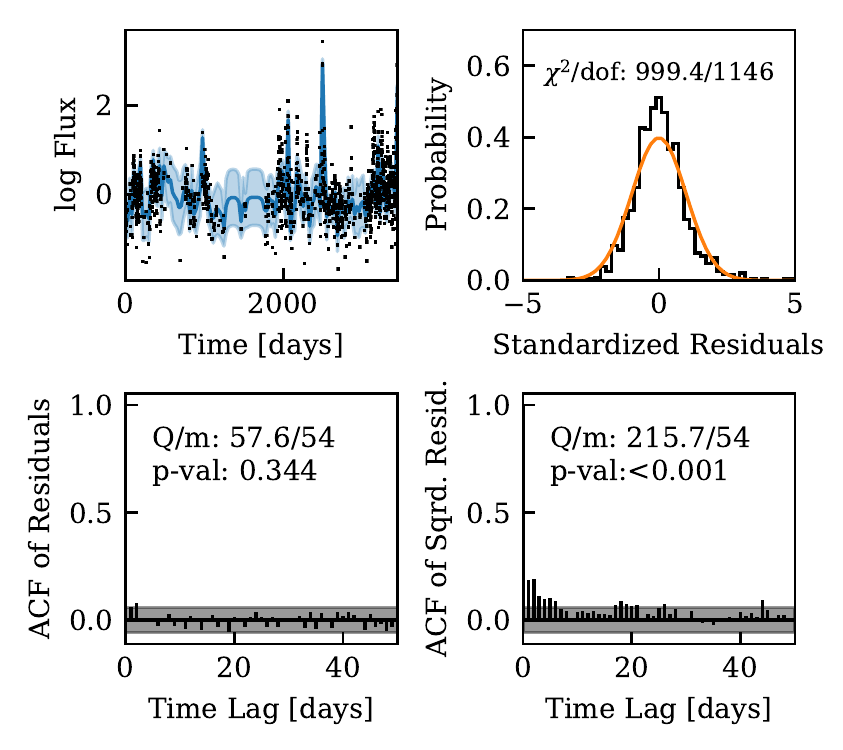}
  \caption{Diagnostic plots for the 3C 279 daily data's CARMA(2,1) model. Top left: Natural logarithm of the light curve (black points) in units of $10^{-6}$ ct cm$^{-2}$ s$^{-1}$. The solid blue line and light blue region denote the MLE light curve and $1\sigma$ error bands based on the best-fitting CARMA(2,1) process. Top right: Histogram of standardized residuals (black), compared with the expected standard normal distribution (orange line). The $\chi^2$/d.o.f. is shown. The small $\chi^2$ value results from the CARMA model capturing some of the variability induced by measurement errors.
  Bottom left: ACF of the standardized residuals compared with the 95\% confidence region assuming a white noise process (gray region). The Ljung-Box $Q$, $m$, and p-value are also shown. The large Ljung-Box p-value suggests that the CARMA model has adequately captured the correlation structure in the light curve. 
  Bottom right: ACF of the squared standardized residuals, with symbols as in the bottom left plot. The low p-value associated with the squared residuals could indicate the presence of nonlinear behavior.}
  \label{fig:assess}
\end{figure}

\begin{figure*}[t]
  \plotone{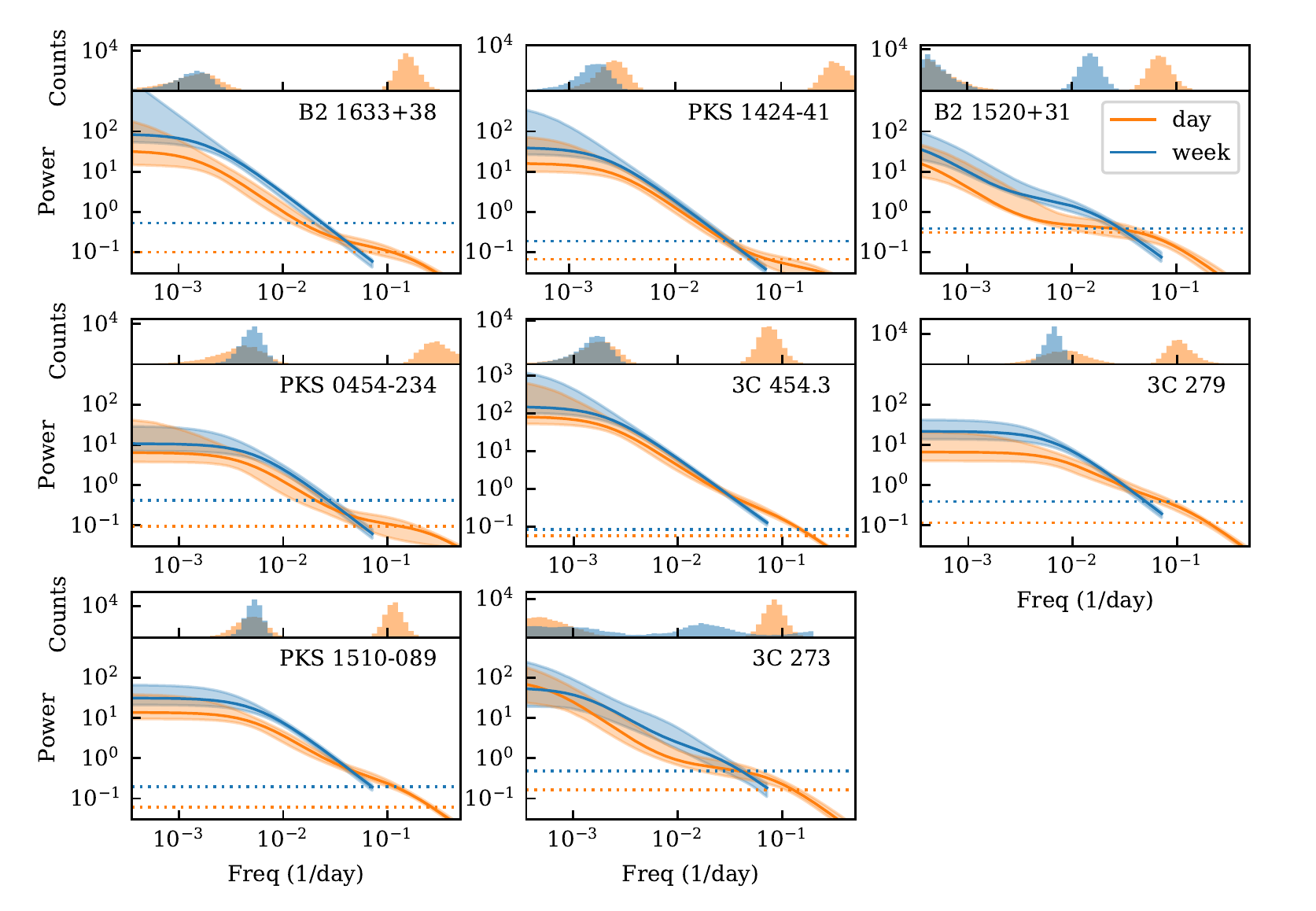}
  \caption{CARMA model PSDs derived from the daily (orange) and weekly (blue) light curves of our sample of FSRQs. Solid lines are the MLE PSDs and shaded regions represent 90\% CIs. PSD power is in units proportional to log(ct cm$^{-2}$ s$^{-1})^2$ / day$^{-1}$. The horizontal dotted orange and blue lines represent the estimated noise levels for the daily and weekly data, respectively. Histograms of the break frequencies derived from the MCMC samples are plotted above each PSD plot. Mismatches between some daily and weekly PSDs may be due to the true PSD having the form of a pure power-law.}
  \label{fig:psd_fsrq}
\end{figure*}

\begin{figure*}[t]
  \plotone{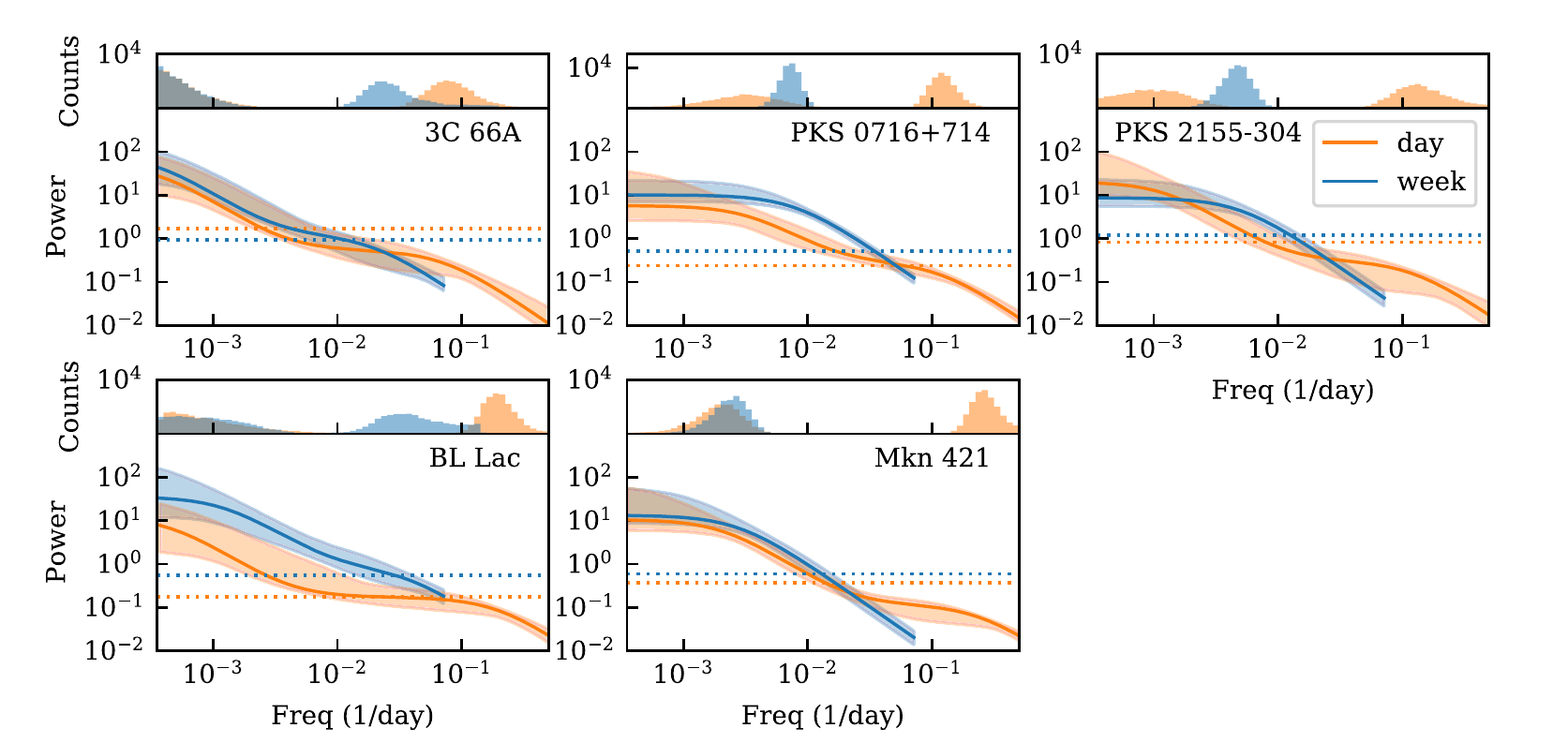}
  \caption{Same as Figure \ref{fig:psd_fsrq}, for the BL Lacs.}
  \label{fig:psd_bllac}
\end{figure*}

We use CARMA modeling to estimate the PSDs of our sample of blazars.
The best-fit CARMA model orders $p$ are presented in Table \ref{table:results}.
The models' $\chi^2$ values and degrees of freedom (d.o.f.) give no p-values less than 0.01, indicating that the CARMA models provide adequate fits to the data.
The autocorrelation functions (ACFs) of the models' standardized residuals are consistent with a white noise process, as well.
In some cases, the $\chi^2$/d.o.f. suggest overfitting.
This is observed in our simulations as well, and is a consequence of the CARMA models capturing some of the white noise behavior induced by measurement errors.
Diagnostic plots for the daily-binned light curve of 3C 279 are shown in Figure \ref{fig:assess}, illustrating some of these points.

PSD estimates are derived from the MCMC sample parameters for each light curve.
Plots of the PSDs with 90\% CIs, accompanied by histograms of the posterior distributions of the break frequencies, are plotted in Figures \ref{fig:psd_fsrq} and \ref{fig:psd_bllac}.
Table \ref{table:results} reports the PSD break timescales with 90\% CIs (99\% CIs in the case of limits), as well as average slopes above the noise level.
Average slopes are reported since, as shown in our simulations, PSDs lacking physical breaks may be simple power laws with slopes equal to the average CARMA PSD slope.
In the table, breaks which are determined to be unphysical are marked with an asterisk.
For the remaining breaks, our methodology provides no evidence that they are artificial.
Of the FSRQs in our sample, a physical break is constrained in PKS 1424-41, PKS 0454-234, and PKS 1510-089, while of the BL Lacs, a physical break is constrained only in Mkn 421.
Additionally, a shorter-timescale break is found in PKS 1510-089 around 9 days.
The average long-timescale break among these FSRQs is $\log  (\tau_L/\text{day}) = 2.55^{+0.34}_{-0.33}$ ($2.87^{+0.41}_{-0.51}$ rest-frame), compared to Mkn 421's break at $2.59^{+0.36}_{-0.14}$ ($2.61^{+0.36}_{-0.14}$ rest-frame).

A significant mismatch exists between the daily and weekly PSDs of B2 1520+31, 3C 273, PKS 0716+714, and PKS 2155-304.
Our simulations reveal that such mismatches can occur when the true PSD is a pure power law.

\section{Discussion} \label{sec:disc}

\subsection{Blazar $\gamma$-ray PSD}
CARMA models adequately describe the logarithms of both the weekly and daily $\gamma$-ray light curves of our 13 blazars, based on residuals between the light curves and the best-fit realization of a CARMA process.
A single stochastic process can therefore account for entire light curves, including both the ``flaring'' and ``quiescent'' states.
ACFs of the models' standardized residuals are consistent with a white noise process, based on the Ljung-Box test, implying that the CARMA models capture the correlation structure of the light curves.
On the other hand, ACFs of the models' squared standardized residuals are often inconsistent with a white noise process (e.g. see Figure \ref{fig:assess}), based on the similar McLeod-Li test \citep{mcl83}, indicating nonlinear behavior.
We therefore cannot conclude that nonlinear behavior is present in the logarithmic blazar $\gamma$-ray light curves.
The DIC show no evidence for significant improvement to fits in CARMA($p,p-1$) models with $p>2$ in any sources, indicating that the PSDs are well described by the sum of one or two Lorentzians in the frequency range of $10^{-3.5}$--$10^{-0.3}$ days$^{-1}$.

\begin{figure}[t]
  \plotone{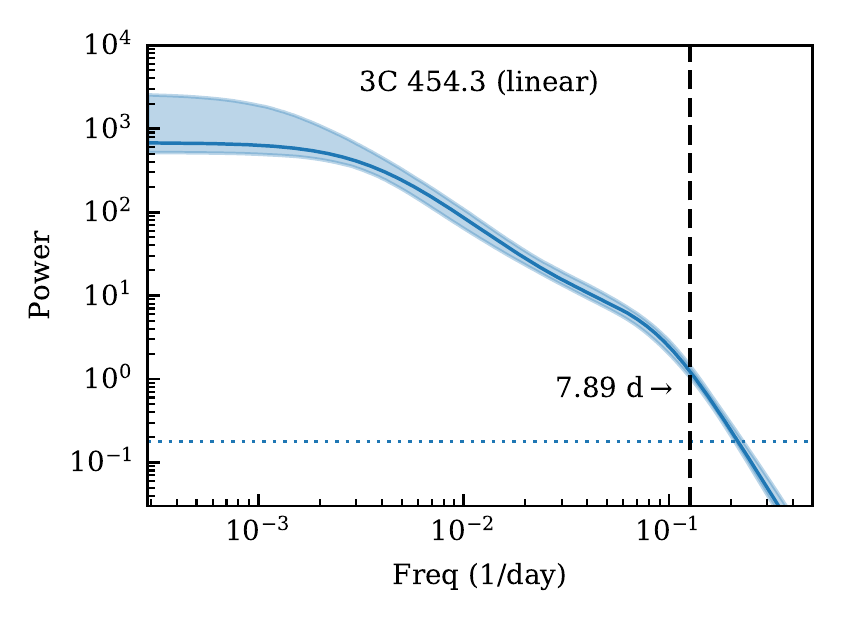}
  \caption{MLE PSD of the CARMA(3,2) model of the daily 3C 454.3 light curve, without using the log of the fluxes.
PSD power is in units proportional to (ct cm$^{-2}$ s$^{-1})^2$ / day$^{-1}$.
The shaded region represents the PSD 90\% CI.
The horizontal dotted line represents the estimated noise level. 
The frequency where \citet{nak13} detect a break is indicated, and is near a break-like feature in the CARMA PSD.}
  \label{fig:3c_linear}
\end{figure}

% breaks
Our analysis revealed the presence of breaks in the PSDs of four sources on timescales $\sim$1 year, and an additional break on a timescale $\sim$ 1 week in a single source, PKS 1510-089.
Previous analyses of the $\gamma$-ray PSDs of the blazars in our sample have also shown evidence for breaks, but at timescales differing from our results: \citet{ack10} and \citet{nak13} detect a break in 3C 454.3 around 6.5 and 7.9 days, respectively, while \citet{sob14} constrain a break in 3C 66A around 25 days and in PKS 2155-304 around 43 days.
It is expected that the longer-timescale breaks we detect are absent from the preceding analyses, since our data set is at least twice theirs in length, and the timescales we find are already close to the maximum recoverable timescale given our methodology.
The absence of our shorter-timescale break in PKS 1510-089 in prior works is less understood, especially since the CARMA(2,1) and sup-OU processes used by \citet{sob14} are fairly similar, but may again be due to the larger amount of data we have access to.
Another possibility is that the adaptive binning algorithm used by \citet{sob14}, which allows higher frequencies to be explored, could distort the sup-OU PSD.
The breaks in 3C 66A and PKS 2155-304 found by \citet{sob14} are present in our CARMA(1,0) PSD models, but not in the CARMA(2,1) PSD models, indicating that they are likely artifacts.
The absence of a short-timescale break in the 3C 454.3 PSD in this work is a consequence of using the logarithm of the light curve, since a similar break is present in the CARMA model of the non-logarithmic fluxes, shown in Figure \ref{fig:3c_linear}.
The break represents a characteristic timescale associated with one or more flaring episodes (e.g. in November 2010), as these portions of the light curve dominate the variability power at high frequencies.
Using the logarithms of the light curves reduces the weight of flares, allowing the PSD to better capture the average light curve behavior.
While \citet{sob14} posit that constraining the high-frequency slope to $-2$ may have led to poor constraints on the characteristic timescales in 3C 454.3, this does not apply to our analysis, as we see that the high-frequency slope in Figure \ref{fig:3c_linear} approaches $\sim-4$.
\citet{kus17} examine light curves spanning a length of time similar to what is used in our analysis, but find no evidence for breaks---this could be a consequence of the PSD estimation technique, which does not aim to model both a break and a high-frequency white noise component.

% slopes
Most blazar PSDs have been found to be consistent with simple power laws \citet{abd10c,nak13,sob14,ram15,kus17}.
In our sample, the CARMA model behavior of B2 1520+31, 3C 279, 3C 273, 3C 66A, PKS 0716+714, PKS 2155-304, and BL Lac resemble what is expected from simple power law PSDs.
In contrast, in the sources where we detect physical breaks, the CARMA models' behavior is inconsistent with our simple power law simulations.
Fitting simple power laws to our CARMA PSDs gives results broadly consistent with those reported in previous studies; although we note that the slope will depend on the frequency range used and how the measurement error-induced white noise component is accounted for.

% QPOs
Although CARMA models can recover QPOs from light curves \citep{kel14}, we find no evidence for QPOs in our sample.
QPOs have been detected in $\gamma$-ray light curves of blazars prior to our analysis \citep{ack15b}, including several blazars in our sample \citep[e.g.][]{san16,san17}.
We did not explore the sensitivity of CARMA to QPOs.
It is possible that the Bayesian model comparison inherent in CARMA modeling, which addresses some of the concerns of false detections put forward by \citet{vau16}, could additionally reduce the method's sensitivity to relatively weak QPOs.

\subsection{Origin of Variability Timescales}
% 1 summary
The PSD provides a statistical characterization of variability that can constrain physical models.
Characteristic timescales in particular can help motivate theories by being compared to physical timescales.
More detailed comparisons can also be made with models of $\gamma$-ray emission for which the PSD can be predicted \citep[e.g.][]{mar14,che16}.

% 2 long timescale
Our results indicate that $\gamma$-ray variability in blazars is characterized by two timescales.
The long-timescale $\tau_L$ represents a maximum correlation time, while the shorter timescale $\tau_S$ represents a minimum variability timescale, or at least implies that variations are smoothed out on timescales shorter than $\tau_S$.
The magnitude of $\tau_L$ makes it more appealing to associate it with variations originating in the accretion disk, where variability can be interpreted in terms of the inner propagating fluctuation models \citep{lyu97}.
Such a model naturally produces a log-normal flux distribution, as in the ``exponential model'' of \citet{utt05}.
Since the jet launching site is in the vicinity of the disk, it is expected that variability in the accretion disk would be imprinted on the jet emission \citep{kat01}.
In this case, $\tau_L$ may be interpreted as one of either the dynamical, thermal, or viscous timescales \citep{cze06}.
Interpretation as a light crossing time is less likely, given the significant amount of correlated variability on shorter timescales.
In some models, timescales associated with viscous instabilities are much longer than our observed values of $\tau_L$ \citep{sie96}, leaving dynamical or thermal timescales as the more likely candidates for the breaks.
The radius implied by the dynamical timescale can be interpreted as the greatest radius at which contributions to the variability are produced; possibly the outer edge of the accretion disk.
The dynamical timescale need not result in periodic variability, as it may drive turbulence in the disk as well \citep{hir09}.
In the scenario where the variability is associated with the disk, the origin of the variability can reasonably be associated with fluctuations caused by disk instabilities \citep{cze06}, or else changes in the mass accretion rate or viscosity parameter $\alpha$ \citep{sha73,lyu97}.

% 3 short timescale
The shorter timescale $\tau_S$ could be associated with the dynamical timescale at a smaller radius, or alternatively with physical processes in the jet itself.
\citet{fin14} model the evolution of electrons in a nonthermal plasma, and calculate the PSDs for both synchrotron self-Compton (SSC) and external Compton (EC) models.
PSD breaks are found to coincide with either light crossing, electron escape, or cooling timescales, and for typical blazar parameters, are all on the order of hours to days.
\citet{che16} expand upon this model by including the effects of particle acceleration and spatial inhomogeneities, although they model only SSC emission, and instead find that the PSD break coincides with the relaxation time of the system.
We note that if the variability originates in the jet, observed timescales will be compressed by the jet's Doppler factor, which is often subject to large uncertainties \citep{lio15}.
The origin of variability in the jet can be ascribed to fluctuations in magnetic field strength, electron injection rate, or Doppler factor \citep[e.g.][]{mas97}.

\begin{figure}[h]
  \plotone{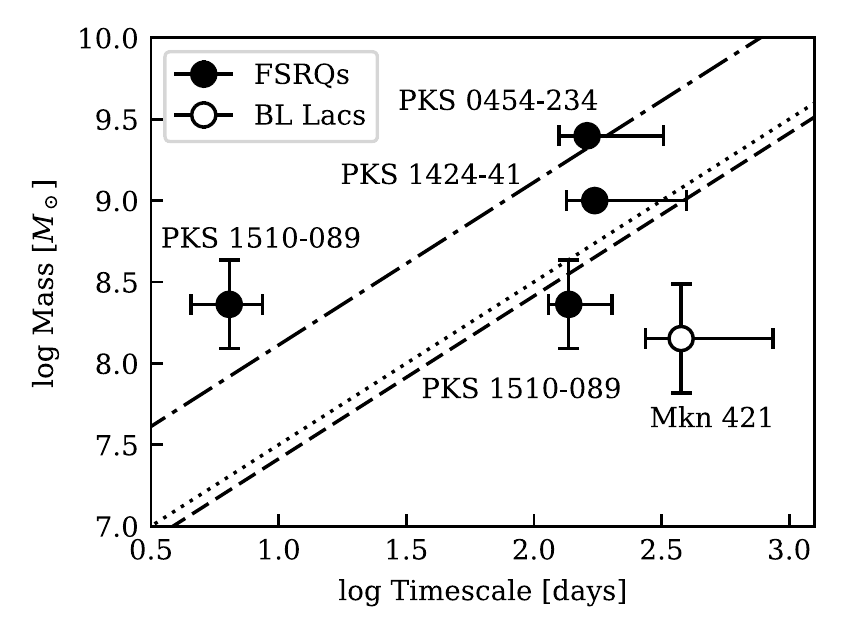}
  \caption{PSD break timescales corrected for cosmological redshifts (i.e. divided by $1+z$), plotted against mass. The FSRQ PKS 1510-089 is plotted twice, corresponding to both $\tau_L$ and $\tau_S$.
Error bars on timescales represent 90\% CIs, while error bars on mass (when present) represent the RMS scatter in values reported in the literature.
Linear relations derived from X-ray variability in a sample of Seyfert 1 galaxies \citep[dotted line;][]{mar03}, and from the GBH Cyg X-1 in its high (dash-dotted line) and low states \citep[dashed line;][]{mch04} are shown.
The long timescales show no apparent correlation with mass.
The short timescale observed in PKS 1510-089 is shorter than predicted by the mass$-$timescale relations which describe non-blazar AGN, and may be due to a higher mass accretion rate, Doppler boosting, or a different physical process than the one causing the X-ray PSD breaks.
  \label{fig:mass}}
\end{figure}

% 4 mass
Correlations between the timescales and independently determined AGN parameters, such as black hole mass $M_\text{BH}$, can help discriminate between the timescales' interpretations.
In the X-ray regime, a linear relation $\tau \propto M_\text{BH}$ has been found for black hole accretion systems ranging from Galactic black hole X-ray binaries to blazars \citep[e.g.][]{mch88,mch04,cha18}.
Such a relation has been found in the optical \citep{kel09,mac10} as well, although \citet{koz17} caution that detections of characteristic timescales close to the light curve length may be dubious.
Our $\gamma$-ray characteristic timescales are plotted against $M_\text{BH}$ in Figure \ref{fig:mass}, along with linear relations derived from X-ray variability in a sample of Seyfert 1 galaxies \citep{mar03}, and from the GBH Cyg X-1 in its high and low states \citep{mch04}.
The long timescales are consistent with the relations derived from X-ray variability, with the exception of Mkn 421, but show no apparent correlation with mass, providing marginal evidence against the longer timescales being dynamical timescales that depend only on mass.
The short timescale observed in PKS 1510-089 is shorter than predicted by the X-ray relations.
Shorter timescales may still be achieved with higher mass accretion rates \citep{mch06}, Doppler boosting (in the case where variability originates within the jet), or different physical processes altogether than those causing the X-ray PSD breaks.

% 5 multi-frequency considerations
Comparing the blazar PSD at different wavelengths, from radio to X-rays, can further constrain emission models.
If the same physical processes are responsible for the variability in different bands, the PSDs will have similar features.
In particular, a relation between $\gamma$-rays and X-rays could imply a common emission region or process for both bands, while a relation between $\gamma$-rays and lower energies (radio through optical) could result from the SSC model for $\gamma$-ray emission, at least in cases where the lower-energy emission is dominated by beamed synchrotron emission.
PSD slopes and break timescales with common origins need not necessarily be identical since frequency-specific variability processes could alter the PSDs at nearby wavelengths in observable ways, such as if the break timescales were relaxation or cooling times, which become larger for lower energies \citep{fin14,che16}.
Alternatively, similar PSDs at different wavelengths might only imply a common origin of the variability, as would be the case in the inner propagating fluctuations model.
PSD breaks on the order of days have been detected in several of our sources in the X-ray \citep{kat01,kat02,mch08,cha18} and the extreme ultraviolet \citep{cag01}.
The $\gamma$-ray PSDs are all close to the measurement noise level at these timescales, however, so detection of similar breaks was not possible.
Many studies find no evidence for breaks in the optical or radio PSDs of the blazars in our sample \citep[e.g.][]{cha08,cha12,ale15,goy17,par17}, although bends on timescales shorter than a day have been found in optical blazar PSDs outside of our sample \citep{ede13,moh16}.

% 6 FSRQ vs BL Lac
It is unclear whether a difference exists between the PSDs of the FSRQs and BL Lacs in our sample, due to the small number of characteristic timescales we could constrain.
However, various models have predicted that such a difference should exist.
If the physical divide between FSRQs and BL Lacs were determined by high and low accretion rates, respectively \citep{ghi09a}, FSRQs would have standard \citet{sha73} accretion disks, while BL Lac accretion disks would operate through a different mechanism, such as advection-dominated accretion flow \citep{nar97}.
In the fluctuating $\alpha$ model of \citet{lyu97}, the characteristic variability timescales differ between these types of disk, with characteristic fluctuation times in advection-dominated disks being shorter.
\citet{ghi09a} also reason that the dominant process responsible for $\gamma$-ray production is EC emission in FSRQs, and SSC emission in BL Lacs.
PSDs associated with EC and SSC are found to exhibit different break frequencies as well as other differing behavior, such as in their low-frequency PSD slopes and the amount of steepening at frequencies higher than the break \citep{fin14}.

% 7 Green's function
An alternative approach to interpreting CARMA models is put forward by \citet{kas17}.
They propose that rather than approximating the PSD, the CARMA process is a direct result of the astrophysics driving the variability.
The CARMA(2,1) model is particularly meaningful, as it has the form of a damped harmonic oscillator.
It may therefore represent increased dynamical complexity in comparison to the CARMA(1,0) model, which describes a damped random walk and has had success describing the light curves of AGN in multiple bands \citep{kel09,mac10}.
The greater complexity of the CARMA(2,1) process appears to only have a physical origin in the case of the daily model of PKS 1510-089.
For PKS 1510-089, we find the damping ratio $\zeta >1$ \citep[see][for notation]{pan83}, meaning that the system is overdamped.
The impulse response, characterized by the Green's function, then has the form of an exponential rise peaking at time $t_\text{max}$ followed by an exponential decay with $e$-folding time $t_\text{e-fold}$ \citep{kas17}.
The MLE of these parameters, with 90\% CIs, are $\zeta=2.60_{-0.43}^{+0.64}$, $t_\text{max}=4.58_{-1.33}^{+1.81}$ days, and $t_\text{e-fold}=39.7_{-13.9}^{+28.7}$ days.
\citet{kas17} also notes that the CARMA(2,1) process' driving noise PSD has the form of white noise, transitioning to violet noise ($\propto f^2$) at high frequencies.
They identify thermal noise within the accretion disk or eddies in a turbulent flow as physical phenomena that can produce a violet noise PSD.

\section{Conclusions} \label{sec:conc}
We present the $\gamma$-ray variability properties of 13 blazars derived from 9.5 years of light curve data from \textit{Fermi}/LAT using the CARMA model fitting technique of \citet{kel14}. 
Through simulations, we find the conditions for which CARMA models produce artificial PSD breaks, and develop a methodology to recognize such breaks using higher-order CARMA models.
We find that CARMA(1,0) models often fail to capture the variability characteristics of the light curves, while CARMA(2,1) models provide adequate descriptions of the variability.
We constrain characteristic break timescales $\sim$1 year in four sources, and an additional timescale $\sim$9 days in a single source, while finding no evidence for QPOs in any source.
The long timescale may represent a thermal or dynamical timescale in the disk, while the short timescale may be associated with the dynamical timescale for a smaller radius, or with processes in the jet.

Determining the physical nature of the $\gamma$-ray variability timescales will allow constraints to be placed on models of blazar emission.
An interpretation of the timescales may be elucidated through expanding variability analyses to larger samples and additional wavelengths.
Incorporating objects with more tightly-constrained black hole masses, or including mass accretion rates and Lorentz factors are also considerations for future studies.

\section{Acknowledgements} \label{acknowledgements}
We thank Brandon Kelly, for his development of and assistance in using CARMA, and Charles Alcock, who offered many constructive comments on an early draft of this paper. We thank the referee for providing helpful suggestions that have improved the paper. A.S. and M.S. were supported by NASA Contract NAS8-03060 (Chandra X-ray Center). M.S. acknowledges the Polish NCN grant OPUS 2014/13/B/ST9/00570.

\software{carma\_pack \citep{kel14}, \; Astropy \citep{astropy}, \; Matplotlib \citep{matplotlib}, \; NumPy \citep{numpy}, \; SciPy \citep{scipy}}

\bibliographystyle{apj}
\bibliography{blazar}
 
 \end{document}